\documentstyle[12pt]{article}
\begin{document}
\def\th{\theta}
\def\eps{\epsilon}
\def\vk{{\bf k}}
\def\de{\Delta}
\def\cj{{\Im}}
\def\kb{k_B\beta}
\def\beq{\begin{equation}}
\def\enq{\end{equation}}
\def\beqn{\begin{eqnarray}}
\def\eenq{\end{eqnarray}}
\def\pl{\parallel}
\baselineskip22pt

\begin{center}
{\large{\bf Superconducting Gap\\ within a Modified Interlayer
            Tunneling Model}}\\
\vspace{0.8cm} 
{\bf Biplab Chattopadhyay$^{\star}$ and A. N. Das$^{\dagger}$}\\
\vspace{0.2cm} 
{Saha Institute of Nuclear Physics, 1/AF Bidhannagar, 
      Calcutta - 700 064, INDIA}  
\end{center} 
\vspace{1cm}

\begin{abstract}

A modified version of the interlayer tunneling model, 
including interlayer single particle hopping (ISPH), is 
considered as a phenomenological model to describe cuprate 
superconductors. The effective ISPH ($t_\perp^{\rm eff}$) 
is taken along with a probability factor $P$, that involves 
the normal state pseudogap ($E_g$). This makes 
$t_\perp^{\rm eff}$ to mimic experimental observations 
that, ISPH is small in the underdoped regime and 
increases towards overdoping. Within the modified model, 
we establish the absence of bilayer splitting as observed 
in case of layered cuprates. Transition temperature 
($T_c$) and  the superconducting gap are calculated. 
A match, to the T-dependent superconducting gap data 
from experiment, is obtained and high values of the 
ratio of the superconducting gap to $T_c$ are recovered.  
Depending on the values of $E_g$, $T_c$ as a function of 
interlayer coupling shows mixed behaviour. This is a 
prediction and can be checked further. 
\end{abstract}
 
\vspace{1ex} 
      
\noindent{PACS numbers: 74.20.-z, 74.62.Dh, 74.25.Bt}     
\vskip18pt
\noindent {\it Keywords}: Interlayer single particle hopping, 
          Bilayer splitting, Superconducting gap.   
\vskip18pt 
\noindent {\large Physics Letters A (to appear).} 
\vfill

\noindent \rule{15.73cm}{0.1mm}\\
{\small $^\star$email: biplab@cmp.saha.ernet.in}; 
{\small $^{\dagger}$email: atin@cmp.saha.ernet.in} 

\newpage 

Surge of research activity on high temperature superconductors (HTS)
in the last few years, have clearly established that, the HTS have
unusual normal as well as superconducting state properties 
different from those of the conventional BCS superconductors.
In the superconducting state, the issues related to the
superconducting gap, such as the symmetry of the gap, ratio of
the gap to $T_c$ and the temperature variation of the gap, are 
worth addressing. There is growing evidence that the gap of the HTS
has $d_{x^2-y^2}$ symmetry with line nodes \cite{dwpap1,dwpap2},
in contrast to
the s-wave gap of the BCS superconductors. Similarly, temperature
dependence of the superconducting gap and the value of the
gap ratio, as observed in experiments \cite{exprat,gapexpt},
are quite different
from those of the conventional superconductors.

Endeavour to explain various unusual properties of HTS,
resulted in the introduction of a number of phenomenological models
which are commonly characterized by a BCS like gap equation together
with well defined quasiparticles in the superconducting state. One 
such model is the {\it interlayer tunneling model} originally 
proposed by P. W. Anderson \cite{pwand}. 
The interlayer tunneling (ILT) model describes bilayer cuprates 
where interlayer single particle hopping (ISPH) is taken to be 
inhibited due to correlation effects, but tunneling of Cooper 
pairs between the layers is considered. This pair tunneling 
amplifies the pairing mechanism within a CuO$_2$ layer and an 
increase in transition temperature ($T_c$) results \cite{grpand}. 
The ILT model could account for high $T_c$ observed in cuprate 
superconductors, but there are difficulties in recovering very 
high values of the gap ratio for realistic parameters as well as 
matching the experimental gap variation data within the 
model. At this point, a relevant question to ask is whether the  
introduction of ISPH necessary. As a matter of fact, in band 
structure calculations, the ISPH matrix elements are found 
to be substantial for bilayer cuprates \cite{band}. In addition, 
experimental observations on sufficiently overdoped cuprates find
metallic-like behaviour of $c$-axis resistivity and the presence 
of Drude peak in the $c$-axis optical conductivity \cite{expres,uchida}. 
These features immediately imply that, in case of HTS the existence 
of ISPH cannot be ruled out at least in the overdoped regime. 

Motivated by the experimental observations \cite{expres,uchida}, 
band structure calculations \cite{band} and the discrepancies
reflected in the properties involving 
the superconducting gap within the ILT model, 
we consider ISPH along with pair hopping to modify the ILT model. 
For underdoped cuprates, the c-axis resistivity shows
semiconducting behavior and Drude peak is absent in the 
optical conductivity along the c-axis. These are signatures 
of the absence or marginal presence of ISPH in the 
underdoped systems and hence, it is considered along with 
a probability factor $P$ such that the effective ISPH is 
very small for underdoping and attains significance 
for overdoped systems. Pair hopping involves the particles 
which are not taking part in the ISPH. Hence, the probability for 
pair hopping is defined \cite{dasil,bcand} as $(1-P)^2$. 

By a detailed study of the density of states within the modified
ILT model, we address the issue of bilayer splitting for layered
cuprates. Obtaining the gap equation within the mean-field
approximation we calculate the superconducting gap as a function
of temperature, and at optimal doping we study the variation of $T_c$  
with the interlayer coupling. Principal results from our calculations 
are as follows. We demonstrate the absence of bilayer splitting 
within the modified model validating the inclusion of ISPH. 
Variation of $T_c$ with interlayer coupling is different  
for different values of $E_g$. Temperature variation of the gap 
from our calculations has an excellant match to the experimental 
data \cite{gapexpt} and ratio of the superconducting gap to 
$T_c$ has high values as in experiments \cite{exprat}.   

\noindent \underline{\it The Model}\,: The Hamiltonian of the coupled 
bilayer complex \cite{dasil,bcand} is written as  
\begin{eqnarray}
H &=& \sum_{i,\vk,\sigma} (\epsilon_\vk - \mu)
    c_{\vk \sigma}^{(i)\dag} c_{\vk \sigma}^{(i)}  
  - \sum_{i,\vk,\vk^\prime} \left[ V_{\vk,\vk^\prime}\, 
    c_{\vk\uparrow}^{(i)\dag} c_{-\vk\downarrow}^{(i)\dag} 
    c_{-\vk^\prime\downarrow}^{(i)} c_{\vk^\prime\uparrow}^{(i)} 
    + h.c \right] \nonumber\\ 
 &-& \sum_{i\neq j,\vk} \left[ T_p^{\rm eff}(\vk)\, 
       c_{\vk\uparrow}^{(i)\dag} c_{-\vk\downarrow}^{(i)\dag} 
       c_{-\vk\downarrow}^{(j)} c_{\vk \uparrow}^{(j)} + h.c \right] 
     + \sum_{i\neq j,\vk,\sigma} \left[ t_\perp^{\rm eff}(\vk)\, 
       c_{\vk \sigma}^{(i)\dag} c_{\vk \sigma}^{(j)} + h.c \right]  
\end{eqnarray}
This is similar to the one considered by Anderson and coworkers 
\cite{gr22and}, barring the last term accounting for the interlayer 
single particle hopping. The operator $c_{\vk\sigma}^{(i)\dag}$ 
($c_{\vk\sigma}^{(i)}$) creates (annihilates) a fermion in the 
$i$-th layer ($i$=1,2) with momentum $\vk$ and spin index 
$\sigma$. Here $V_{\vk,\vk^\prime}$ is the interaction potential 
forming Cooper pairs in the $ab$-plane and $\mu$ is the chemical 
potential. For the $ab$-plane band dispersion $\epsilon_k$, we use
the realistic band structure obtained from a six parameter tight
binding fit $[t_0, t_1, t_2, t_3, t_4, t_5]$
= [0.131,-0.149,0.041,-0.013,-0.014,0.013]\,eV to the ARPES data
\cite{norman} on Bi2212. This six parameter band dispersion,
used elsewhere \cite{biplab}, shows flat bands in the 
Brillouin zone and the corresponding density of states (DOS)  
has a power law singularity known as extended van Hove 
singularity. This is characteristic to the high-$T_c$ 
cuprates \cite{evhs}. The effective ISPH matrix element 
is taken to be $t_\perp^{\rm eff}(\vk) = t_\perp^b(\vk) P$, 
where $t_\perp^b(\vk)=t_\perp((\cos k_xa-\cos k_ya)/2)^2$ 
is the $\vk$-dependent ISPH adapted from the band 
structure calculations \cite{band} with $t_\perp$ being the 
bare ISPH matrix element ($a$ is the lattice constant) and 
$P$ is a probability factor determined by the normal state 
pseudogap found recently in layered cuprates \cite{psgap}. 
Particles that are not taking part in the ISPH, are available 
for pair tunneling and complementary probability for each partner 
of a pair is ($1-P$). This 
fixes the tunneling probability of a pair as $(1-P)^2$, and 
following Anderson \cite{grpand} we take the effective pair 
tunneling $T_p^{\rm eff}(\vk) = [(t_\perp^b(\vk))^2/|t_1|]\,(1-P)^2$, 
where $t_1$ is the nearest neighbour hopping matrix element 
of the $ab$-plane band dispersion \cite{norman}. 

\noindent \underline{\it Interlayer Single Particle Hopping 
Probability and the Normal State Pseudogap}\,: In experiments on 
layered materials, an important measurable quantity is the 
$c$-axis resistivity which, when studied as a function of doping 
($\delta$), provides usefull information about the strength of 
the effective single particle hopping between the layers. For 
layered cuprates, the $c$-axis resistivity shows semiconducting 
behaviour in the underdoped regime \cite{infra} and has 
metallic-like temperature dependence in the heavily overdoped 
regime \cite{expres,uchida}. This has direct bearing on the ISPH and  
implies that the effective ISPH ($t_\perp^{\rm eff}$) is strong 
enough for overdoping, decreases gradually through optimal 
doping and becomes weaker towards underdoping. The probability  
factor $P$, buried in $t_\perp^{\rm eff}$ could account for 
this doping dependence, when suitably expressed in terms of 
the temperature and the normal state pseudogap $E_g$ of cuprate 
superconductors. The magnitude of $E_g$ is found to be 
vanishingly small for overdoping and it increases towards 
underdoping \cite{psgvar}. Previously, we considered an exponential
form \cite{dasil,bcand} for the probability factor, $P=\exp(-E_g/T)$ 
and found that the experimental gap variation data, in the low
temperature region, were not precisely recoverable. We also
noticed that a linear $T$-dependent form for $P$ could significantly
improve the matching of experimental data. Thus, here we consider a
linear (low-temperature) $T$-dependent form (for $E_g\gg T$) of the
probability factor as 
\begin{equation}
P = {{T}\over{E_g + T {\rm e}^{-{E_g}/T}}} 
\end{equation} 
where $T=\beta^{-1}$ with the Boltzmann constant $k_B$ set to 
unity. Clearly, $P\rightarrow 1$ for $E_g\rightarrow 0$ 
(heavily overdoped situation) and $P\rightarrow 0$ as 
$E_g\rightarrow \infty$ (underdoped situation). Thus 
$P$, as in Eq(2), consistently mimics the observable doping 
dependence of the effective ISPH. 
Though a concrete justification of Eq(2) is not possible at present,
we present plausible arguments in favour of the linear 
$T$-dependence of $P$ in connection with the RVB model. 

Within the RVB framework, where spin-charge separation 
yields spinon and holon 
quasiparticles, Anderson argued that a real hole can come 
into being in a layer only when a holon combines with a 
spinon and this hole can then hop to another layer. Thus, 
within RVB model, $c$-axis transport is proportional to the 
spinon density. Since, in the RVB model, the spinon density  
is proportional to $T$, hence the linear $T$-dependence of  
$c$-axis transport follows. In calculations, we incorporate the 
momentum dependence of the normal state pseudogap, which is found to 
be of $d_{x^2-y^2}$ symmetry \cite{psgap}, and we take 
$E_g(\vk)=E_g|\cos k_xa -\cos k_ya|$.  

\noindent \underline{\it Self Consistent Equations for the Superconducting 
Gap and the Chemical Potential}\,:\\ Within the Hartree-Fock 
approximation, self consistent equations for the superconducting 
gap $\Delta_\vk$ and the chemical potential $\mu$ (fixing the 
average number of electrons per site $n=1-\delta$) are derived 
for the modified model under consideration. Decoupling of 
the four fermion terms in the Hamiltonian of Eq(1) gives 

\begin{eqnarray}
H &=& \sum_{i,\vk,\sigma} (\epsilon_k - \mu) 
      c_{\vk \sigma}^{(i)\dag} c_{k \sigma}^{(i)} - 
      \sum_{i,\vk} \left[\Delta_{\vk}\, c_{\vk \uparrow}^{(i)\dag} 
      c_{-\vk \downarrow}^{(i)\dag} + h.c \right] \nonumber\\
  &+& \sum_{i\neq j,\vk,\sigma} \left[t_\perp^{\rm eff}(\vk)\, 
      c_{\vk \sigma}^{(i)\dag} c_{\vk \sigma}^{(j)} + h.c \right] 
\end{eqnarray}
where the $\vk$-dependent superconducting gap is  
\begin{equation}
{\Delta}_\vk = \Delta_{i,\vk} = \sum_{\vk^\prime} 
  V_{\vk,\vk^\prime}\, \langle c_{-\vk^\prime \downarrow}^{(i)} 
  c_{\vk^\prime \uparrow}^{(i)}\rangle  
  + T_p^{\rm eff}(\vk)\, \langle c_{-\vk \downarrow}^{(j)} 
  c_{\vk \uparrow}^{(j)}\rangle 
\end{equation}
which includes pairing in one layer as well as a contribution from 
pairing in the other layer owing to the pair tunneling mechanism. Since 
the in-plane pairing is identical in both the layers, we suppress 
labelling the gap parameter by layer index. Coherent single particle 
hopping between the layers produces two quasiparticle bands 
$$
E_\vk^- =\sqrt{\{\epsilon_\vk - \mu-t_\perp^{\rm eff}(\vk)\}^2 
+ \Delta_\vk^2}\eqno(5a) 
$$ 
and 
$$
E_\vk^+ =\sqrt{\{\epsilon_\vk - \mu+t_\perp^{\rm eff}(\vk)\}^2 
+ \Delta_\vk^2}\eqno(5b)
$$  
\addtocounter{equation}{1}
and the gap parameter turns out to be 
\begin{equation}
\Delta_\vk = \frac{\displaystyle\sum_{\vk^\prime} \Delta_{\vk^\prime}\, 
		V_{\vk,\vk^\prime} 
		\left(\chi(E_\vk^-)+\chi(E_\vk^+)\right)\!\!\bigg/2}
	       {1 - T_p^{\rm eff}(\vk) 
		\left(\chi(E_\vk^-)+\chi(E_\vk^+)\right)\!\!\bigg/2} 
\end{equation}
where $\chi(E_\vk^\pm)=\frac{1}{2E_\vk^\pm}
\tanh\left(\frac{\beta E_\vk^\pm}{2}\right)$. The pairing 
potential can be expanded as 
$V_{\vk,\vk^\prime} = V\,\eta_\vk\,\eta_{\vk^\prime}$, where 
$\eta_k$ is the basis function of the $c_{4v}$ point group. 
This makes it possible to write $\Delta_\vk$ in terms of a 
$\vk$-independent gap $\Delta$, the basis function $\eta_\vk$ 
and the interaction parameter $V$. Finally, the self consistent 
equation for superconducting gap becomes  
\begin{equation}
{1\over{4V}} = \frac{1}{N} \sum_{\vk} 
	     \frac{\eta_\vk^2\,\left(\chi(E_\vk^-) + \chi(E_\vk^+)\right)\!\!\bigg/2} 
	     {1 - T_p^{\rm eff}(\vk)\left(\chi(E_\vk^-) + \chi(E_\vk^+)\right)\!\!\bigg/2} 
\end{equation}
and the expression for chemical potential is  
\begin{equation}
1-\delta = 1-\frac{1}{N} \sum_\vk \left(\epsilon_\vk-\mu -t_\perp^{\rm eff}\right) 
	   \chi(E_\vk^-)
	 - \frac{1}{N} \sum_\vk \left(\epsilon_\vk-\mu +t_\perp^{\rm eff}\right) 
	   \chi(E_\vk^+)
\end{equation}
									 
Regarding the order parameter symmetry, though the issue is yet to 
be settled on a definite footing, a consensus seems emerging in the 
many experiments done on layered cuprates. The angle resolved  
photoemission spectroscopy (ARPES) measurements and other phase 
sensitive experiments indicate the $d_{x^2-y^2}$ symmetry of the 
order parameter \cite{dwpap1,dwpap2}, and we consider the same in 
our calculations. This means the basis function 
$\eta_\vk = (\cos k_xa -\cos k_ya)/2$, which changes sign under 
$\pi/2$ rotation. The interaction parameter $V$ is thus the nearest 
neighbour and attractive. 

\noindent \underline{\it Bilayer Splitting}\,: For bilayer cuprates,
the presence of effective single particle
hopping between the layers should be observable in the single 
particle density of states (DOS). In our calculations, the  
realistic six parameter dispersion for Bi2212, used to model 
the in-plane band structure, shows a power law singularity in 
the DOS in the form of a single peak. Inclusion of ISPH between 
the CuO$_2$ layers would split the bands and the corresponding 
DOS will have two peaks. This phenomenon is known as 
{\it bilayer splitting}. However, no splitting is observed 
experimentally in the 
Bi2212 systems \cite{bisplit} even at low temperatures where 
broadening due to finite lifetime of quasiparticles is expected 
to be small. In Fig.1, we plot the DOS, $N(\xi)$ at two different 
temperatures where $\xi_\vk = \eps_\vk \pm t_\perp^{\rm eff}(\vk)$.
Temperature dependence of $N(\xi)$ comes only through the 
probability factor $P$. Parameters for the plots are chosen to 
be $t_\perp = 40\,meV$ and $E_g = 8\,meV$, same as used to match 
the gap-variation data in Fig.3. At low temperatures $T=10\,K$ and 
with no broadening (i.e. broadening parameter $\Gamma=0$),  
we find that $N(\xi)$ has two peaks (solid line) 
separated by an energy $\sim 3\,meV$. However, inclusion of a 
small broadening $\Gamma=3\,meV$ (dashed line) smears the two 
peaks and one broad peak appears. At a higher temperature 
$T=50\,K$, the larger peak separation energy $\sim 21\,meV$ is 
due to the enhanced probability of the interlayer single particle 
hopping. Inclusion of nonzero $\Gamma$ (dashed lines) broadens 
the peaks and the value of $\Gamma$ needed for complete 
smearing of two peaks is $\Gamma \geq 10\,meV$. This is 
demonstrated in the lower panel ($T=50\,K$) of Fig.1. 

It should be mentioned here that, the temperature dependence 
of the broadening parameter for layered cuprates is noted to 
follow \cite{exprat} the relation 
$\Gamma\,(meV) \approx \Gamma_0 + 2.5 T_c (T/T_c)^3$, 
where $\Gamma_0$ is the zero temperature value. 
Although high values of $\Gamma_0\sim 15\,meV$ is observed for 
Bi-cuprates \cite{exprat}, in our case a $\Gamma_0\sim 7-8\,meV$ 
could be enough to remove the splitting at all temperatures. 
Moreover, at low temperatures ($T=10\,K$) the peak separation 
is small $\sim 3\,meV$ (solid line for $T=10\,K$ in Fig.1), 
whereas in the ARPES experiments the energy resolution is far 
beyond this level (resolution function FWHM$\sim 19\,meV$)
\cite{resolu} and the small peak separation would probably 
remain unresolved.  Thus, in the cleanest sample and at 
very low temperature, even if one ideally takes 
$\Gamma_0\rightarrow 0$, the bilayer splitting in 
Bi-cuprates would remain unobservable by experiments. 
Finally, it has been noted that YBCO systems have a smaller 
$\Gamma_0\sim 1.5\,meV$ \cite{exprat} and also have 
orthorhombic distortions, which might explain why 
splitting could possibly be observed in YBCO \cite{ybco} 
in contrast to the Bi-cuprates. 

So far, we have discussed the issue of bilayer splitting by 
calculating the normal state density of states. But, 
in actual ARPES experiments, bilayer splitting is 
addressed by studying the photoemission intensity curves in the 
superconducting state. Within the modified ILT model, we have 
calculated the photoemission intensity curves \cite{bcand} in the 
superconducting state and found that, with an energy resolution 
much better than that in actual experiments \cite{resolu} and 
with a very small boradening $\Gamma \sim 1\, meV$, bilayer 
splitting remains unobservable for temperatures ranging upto 
$50\, K$. In other words, the inclusion of an effective ISPH 
in the modified ILT model is in perfect tune with the 
experimental observations.  
 
\noindent \underline{\it Issues related to the Superconducting Gap}\,:
The superconducting gap and other quantities are calculated by 
simultaneous solutions of the self consistent  
Eq(7) and Eq(8) for the gap and the chemical potential respectively, 
implementing numerical techniques.  

In our calculations,
doping is kept fixed at the optimum level ($\delta_{\rm opt}$) at 
which $T_c$ attains its maximum value $T_c^m$. Position of 
$\delta_{\rm opt}$ is found to have a slow dependence on the 
value of $E_g$ considered. A study of $T_c$ versus $\delta$, 
for different $E_g$, shows that with $E_g\leq 5\,meV$, $T_c$ has  
a two-peak structure, whereas for higher $E_g$ a one-peak bell-shaped 
form is recovered consistent with that in layered cuprates. Thus, 
we enforce the condition $E_g>5\,meV$. Value of bare single particle 
hopping matrix element is taken to be $t_\perp=40\,meV$ in 
accordance with the band structure calculations suggesting  
$t_\perp/|t_1|\sim 0.2-0.3$. Interaction parameter is taken to 
be $V=80\,meV$ which makes $T_c^m\sim 100\,K$ for $E_g\sim T_c^m$. 
This value of $V$ is kept fixed throughout in this communication. 

In Fig.2(a) we plot $T_c^m$ as a function of $t_\perp$. Different 
curves, labelled by capital alphabets, correspond to various $E_g$ 
values as, A:6, B:10, C:15 and D:20 in $meV$. Clearly, variation 
of $T_c^m$ with $t_\perp$ may take different shapes depending on 
the values of $E_g$. For small $E_g$, 
the suppression of coherent single particle hopping between the 
layers is weak and $T_c^m$ decreases with $t_\perp$, whereas 
large $E_g$ makes ISPH less probable and in turn the pair tunneling 
probability gets enhanced, which results a rise in $T_c^m$  
with $t_\perp$. In the Anderson limit ($E_g=\infty$), $T_c^m$ 
grows rapidly with $t_\perp$. 

Application of pressure on a sample along $c$-axis could reduce 
the out-of-plane lattice constant which results in an increase  
of the interlayer coupling. In the plot of $T_c^m$ versus 
$t_\perp$ in Fig.2(a), we find that, for large $E_g$ 
(underdoped materials), $T_c^m$ increases with $t_\perp$, 
for small $E_g$ (overdoped materials), $T_c^m$ decreases with 
$t_\perp$ and at some intermediate $E_g$, $T_c^m$ remains 
unchanged. Thus, within our model, $c$-axis applied pressure 
would result, an increase of $T_c$ for underdoping, a 
decrease of $T_c$ for overdoping and an unaffected $T_c$ for 
some intermediate doping. Similar varied behaviours of $T_c$ 
are seen for differently doped cuprates under hydrostatic or 
uniaxial pressure \cite{pressu}. However, it may be mentioned 
that hydrostatic pressure could also inflict changes in the 
$in~plane$ parameters. Hence, we leave this result of 
various $T_c^m$ variation with $t_\perp$ as a prediction 
which could be checked by further experiments.  

Variation of the zero temperature superconducting gap 
($\Delta_\vk^{max}(0)$) as a function of $t_\perp$ is shown 
in Fig.2(b). The $E_g$ values corresponding to different curves 
are same as used in Fig.2(a). Here, irrespective of the values 
of $E_g$, $\Delta_\vk^{max}(0)$ increases with $t_\perp$ in the 
same fashion for every curve. 

A plot of $T_c^m$ as a function 
of $E_g$, for different $t_\perp$, is given in Fig.3. As is 
obvious, $T_c^m$ increases with increasing $E_g$ due to the 
progressive enhancement of effective pair tunneling probability, 
caused by $E_g$. As is seen, at some $E_g$ close to $\sim 12\,meV$, 
$T_c^m$ for all the curves are same, implying that the transition 
temperature is independent of $t_\perp$. In fact, we crosscheck 
that, $T_c^m$ as a function of $t_\perp$, for $E_g\approx 12\,mev$, 
is nearly flat.  

Next, we come to the issue of the temperature dependence of the 
$d$-wave superconducting gap. A plot of scaled supercondcting 
gap $\Delta_\vk^{max}(T)/\Delta_\vk^{max}(0)$ as a function 
of reduced temperature $T/T_c$ is given in Fig.4. Solid line 
is the BCS-form of gap-variation and solid squares are 
experimental data on Bi-cuprate \cite{gapexpt}. 
Except at the edges ($T/T_c = 0\,\&\,1$) the   
locus of the experimental data is well below the BCS-curve and 
the absence of low temperature saturation in the experimental data 
stands out as an anomalous characteristic of high-$T_c$ cuprates. 
The dashed lines are from our calculations, but, the one 
very close to the BCS-curve represents the gap variation for 
the original ILT model ($E_g=\infty$). The dashed curve with 
$E_g=8\,meV$ is within the modified ILT model which matches 
experimental data very well. To the best of our 
knowledge, this is the first ever calculations where precise 
matching of experimental gap-variation data is obtained. An 
important and related quantity is the ratio of the zero temperature 
superconducting gap to $T_c$. In experiments on layered high-$T_c$ 
cuprates, high values of $2\Delta_\vk^{max}(0)/T_c \sim 5-7$ 
is observed \cite{exprat}, compared to the 
BCS superconductors where it is $\sim 3.5$. In the inset of Fig.4 
we plot the gap-ratio as a function of $t_\perp$ for the same 
parameters used to fit the experimental gap-variation data. The 
gap-ratio increases with $t_\perp$ and within the realistic range 
of $t_\perp$ ($\sim 30-45\,meV$) we recover the ratio $\sim 5-7$. 
To be precise, at $t_\perp=40\,meV$, we get  
$2\Delta_\vk^{max}(0)/T_c =5.8$. Note that, for same parameters 
and with $E_g=\infty$ (Anderson limit), we find 
$2\Delta_\vk^{max}(0)/T_c = 4.4$. High value of the gap-ratio 
and its increasing trend with $t_\perp$ within the modified model, 
could be understood as follows. 

For any finite $E_g$, when $T\rightarrow 0$, the probability 
factor $P\rightarrow 0$, ISPH is completely blocked and 
$\Delta_\vk^{max}(0)$ grows with $t_\perp$ since $t_\perp$ 
acts only to enhance $T_p^{\rm eff}$. But, for any finite 
$T>0$ the probability factor $P\neq 0$, blocking of the effective 
ISPH is not complete and one gets a transition temperature 
$T_c(P\neq 0) < T_c(P=0)$ resulting in a high value of the 
gap-ratio. Regarding the increasing trend of gap-ratio, consider 
the situation when $T_c^m$ decreases or saturates with $t_\perp$ 
for small $E_g$ (as in Fig.2(a)). But notice that, 
$\Delta_\vk^{max}(0)$ always rises with $t_\perp$ (see Fig.2(b)) 
yielding the observed result of the gap-ratio. 
Even for the case when $T_c$ increases with $t_\perp$ (for high 
$E_g$), note that its increase is slower than 
$\Delta_\vk^{max}(0)$ which effectively increases the gap-ratio 
with $t_\perp$. For a fixed value of $t_\perp$, one gets an 
increasing gap-ratio with decreasing $E_g$ which could also 
follow from similar arguments. 

To conclude, we have considered a modified interlayer pair
tunneling model where pair tunneling is accompanied by 
interlayer single particle hopping. Within the model, absence 
of bilayer splitting is established corresponding to 
experimental situation. Different types of variation 
of $T_c$ as a function of interlayer coupling, for different 
values of the pseudogap, comes out as a prediction that  
corresponds to $T_c$ variation under $c$-axis pressure. 
Temperature dependent superconducting gap from our calculations 
correctly reproduces the experimental data, and for realistic 
parameters, high values of the gap-ratio are obtained.

\newpage

\noindent{\Large{\bf Figure captions:}}  
\vskip 0.3 cm

\noindent 
Fig.1. Single particle density of states $N(\xi)$ is plotted at two 
  different temperatures $T=10\,K$ and $T=50\,K$. Values of broadening 
  parameter $\Gamma$ used are given in the figures. The full bandwidth 
  of energy $\xi$ is not shown, since the peaks in the DOS are only 
  of concern for the purpose of addressing the issue of bilayer 
  splitting.  
\vskip 0.5 cm

\noindent
Fig.2. (a) Mean-field transition temperature ($T_c^m$) at optimal 
  doping as a function of bare interlayer coupling $t_\perp$ for 
  different $E_g$ represented by alphabetic labels as A:6, B:10, 
  C:15, D:20 in $meV$. The interaction parameter is $V=80\,meV$.\\
  (b) A plot of the maximum value of the zero temperature 
  superconducting gap. Values of $E_g$ for different 
  curves are as shown and the alphabetic labels corresponds to same 
  $E_g$ values as in (a).  
\vskip 0.5 cm

\noindent
Fig.3. Transition temperature as a function of the normal state pseudogap 
  magnitude $E_g$ for three different values of $t_\perp$, as shown 
  in the figure. Doping is set to the optimum level (for $E_g=8\,meV$) 
  and $V=80\,meV$. 
\vskip 0.5 cm

\noindent
Fig.4. Temperature variation of the scaled superconducting gap   
  ($\Delta_\vk^{max}(T)/\Delta_\vk^{max}(0)$) as a function of the  
  reduced temperature ($T/T_c$) is presented. Solid line corresponds 
  to the conventional BCS superconductors and solid squares are 
  experimental data from Ref.\cite{gapexpt}. Dashed lines are from 
  our calculations for different $E_g$ shown. 
  [Inset: A plot of the ratio $\Delta_\vk^{max}(0)/T_c$ as a function 
  of $t_\perp$. The parameters are $V=80\,meV$ and $E_g=8\,meV$ that 
  yielded match to the experimental gap-variation data as in the main 
  figure.]   
\vskip 0.5 cm

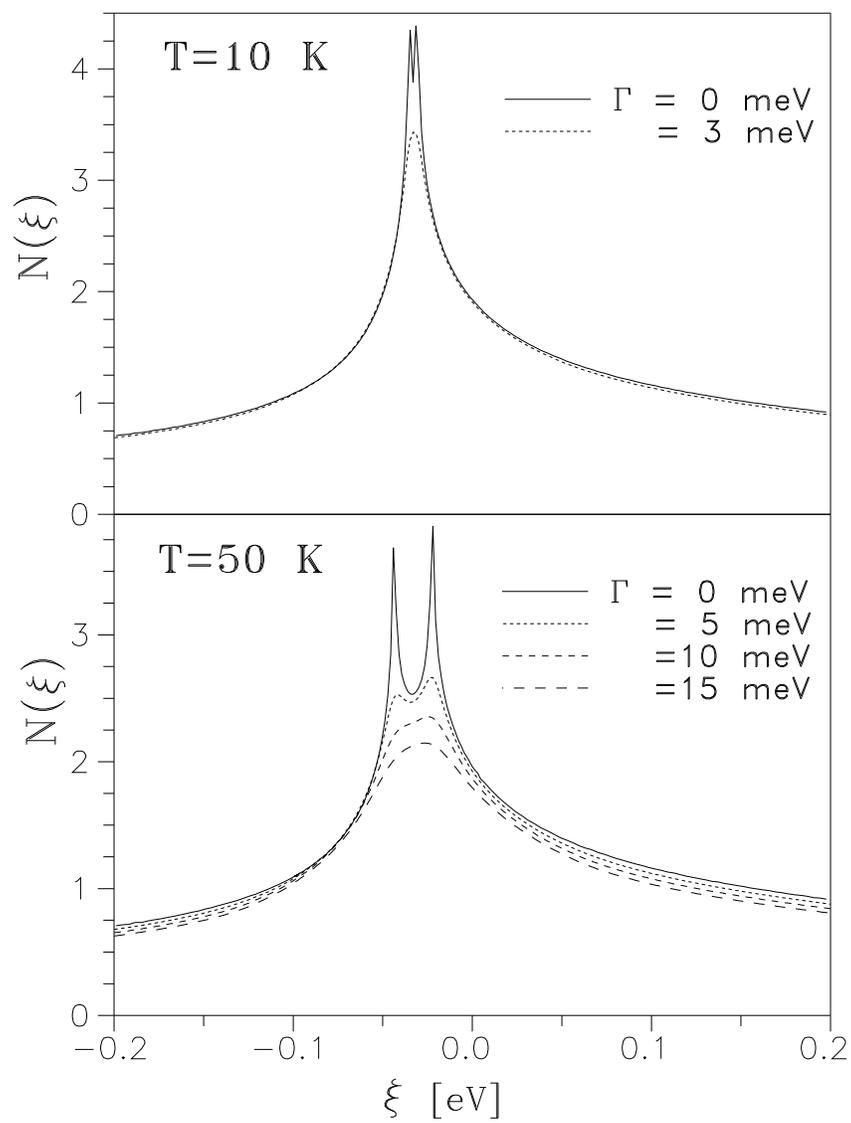
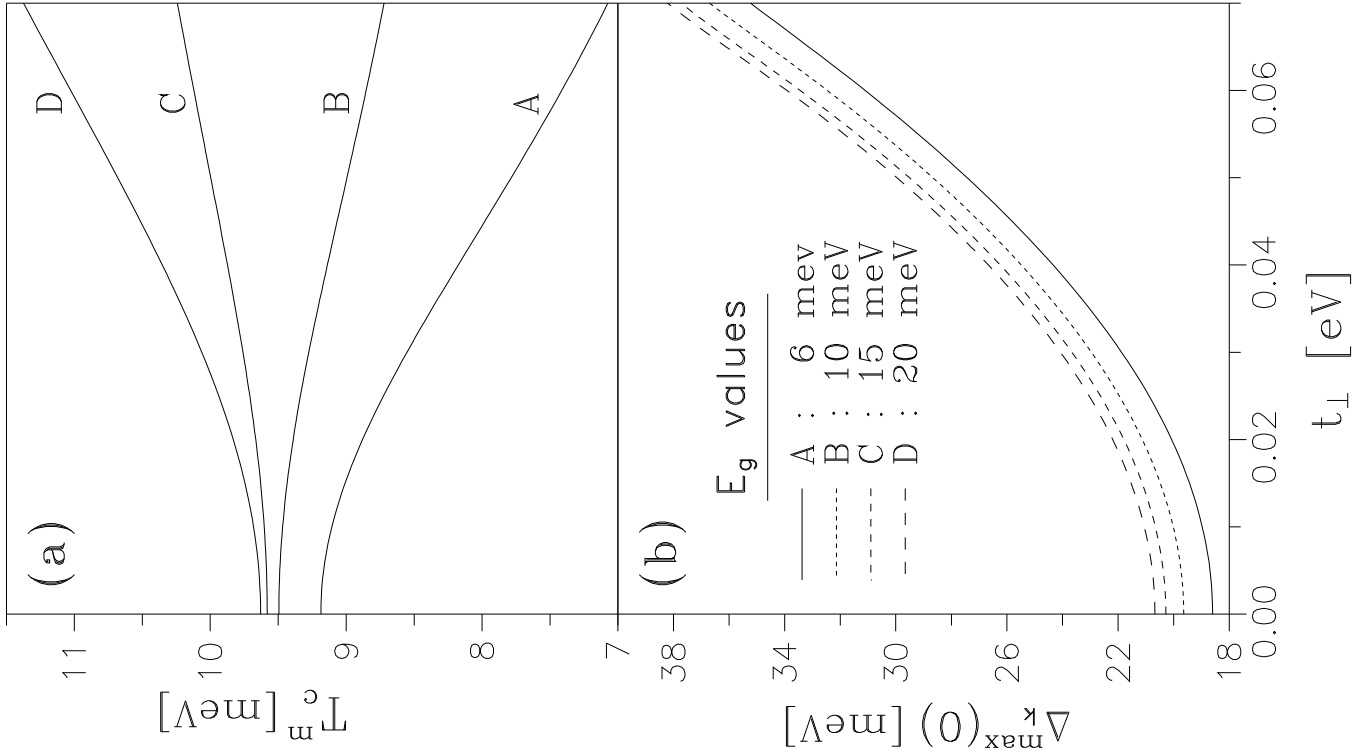
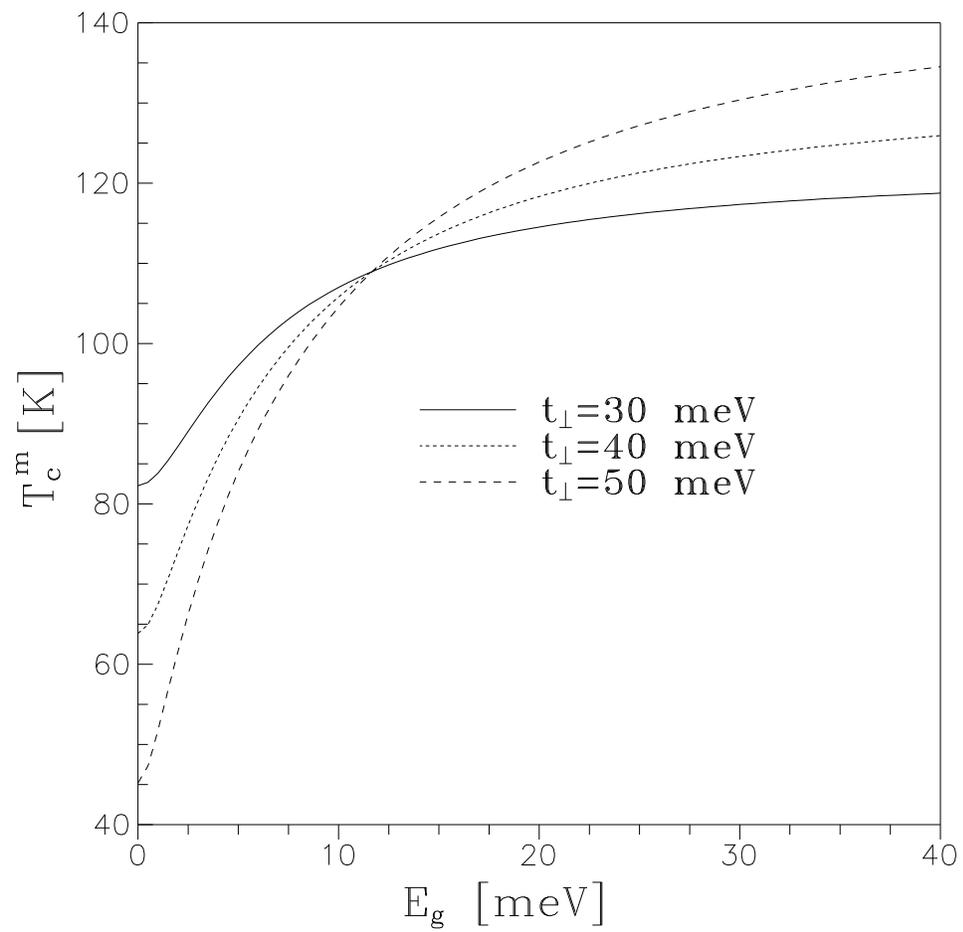
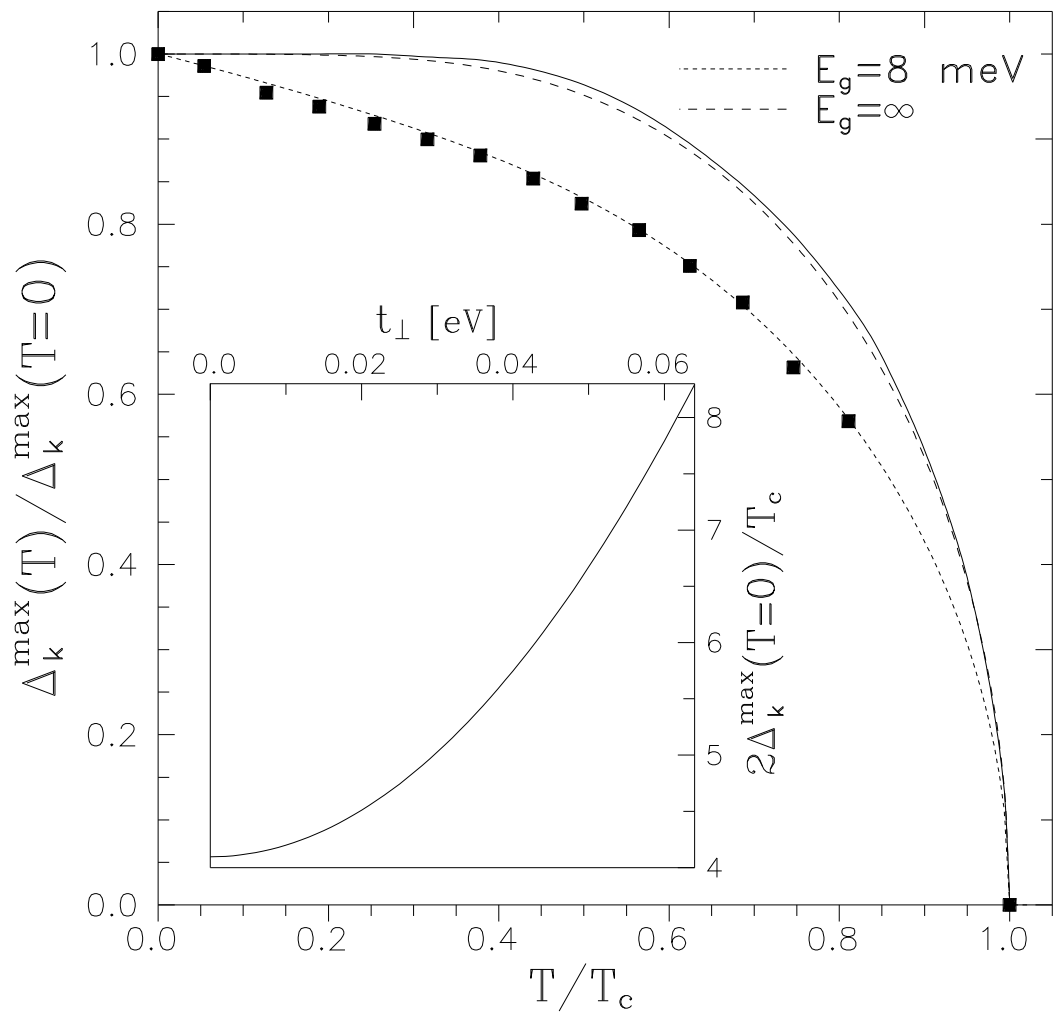
\end{document}